\documentclass{sig-alternate}

\usepackage{epstopdf}
\usepackage{amsmath}
\usepackage{amssymb}
\usepackage{subfigure}
\usepackage{color}
\usepackage{textcomp}
\usepackage{mathtools}
\usepackage{url}
\usepackage{mathrsfs}
\usepackage{times}

\usepackage[ruled,noend,linesnumbered]{algorithm2e}
\usepackage{tikz}
\usetikzlibrary{calc}
\usetikzlibrary{positioning}

\allowdisplaybreaks[1]

\newcommand{\inlheader}[1]{\vspace{1.5mm}\noindent{\bf #1.\ }}
\newcommand{\matr}[1]{\mathbf{#1}}
\newcommand{\set}[1]{\mathcal{#1}}
\newcommand{\algrole}{{\sc NetLand }}
\renewcommand{\vec}[1]{\mathbf{#1}}
\DeclareMathOperator{\sign}{sign}

\DeclareMathOperator{\proj}{proj}
\DeclareMathOperator{\reduce}{reduce}
\DeclareMathOperator*{\argmax}{arg\,max}
\DeclareMathOperator{\mask}{mask}

\newtheorem{defn}{Definition}

\newenvironment{citem}{
\begin{itemize}
  \setlength{\itemsep}{3pt}
  \setlength{\parskip}{0pt}
  \setlength{\parsep}{0pt}
}{\end{itemize}}

\newcommand{\mylistbegin}{
	\begin{list}{$\bullet$}
		{
			\setlength{\itemsep}{-2pt}
			\setlength{\topsep}{1pt}
			\setlength{\leftmargin}{1em}
			\setlength{\labelwidth}{1em}
			\setlength{\labelsep}{0.5em} } }
	
	\newcommand{\mylistend}{
	\end{list}  }

\begin{document}
\title{Network Cartography: Seeing the Forest and the Trees}
\numberofauthors{1}
\author{
	\alignauthor
	Jia Wang,\quad Kevin Chen-Chuan Chang,\quad  Hari Sundaram\\
	\affaddr{Department of Computer Science}\\
	\affaddr{University of Illinois at Urbana-Champaign}\\
	\affaddr{Champaign, IL 61820}\\
	\email{\{jiawang4, kcchang, hs1\}@illinois.edu}
}

\maketitle

\begin{abstract}
\label{abstract}
Real-world networks are often complex and large with millions of nodes, posing a great challenge for analysts to quickly see the big picture for more productive subsequent analysis. We aim at facilitating exploration of node-attributed networks by creating representations with conciseness, expressiveness, interpretability, and multi-resolution views. We develop such a representation as a {\it map} --- among the first to explore principled network cartography for general networks. In parallel with common maps, ours  has landmarks, which aggregate nodes homogeneous in their traits and interactions with nodes elsewhere, and roads, which represent the interactions between the landmarks. We capture such homogeneity by the similar roles the nodes played. Next, to concretely model the landmarks, we propose a probabilistic generative model of networks with roles as latent factors. Furthermore, to enable interactive zooming, we formulate novel model-based constrained optimization. Then, we design efficient linear-time algorithms for the optimizations. Experiments using real-world and synthetic networks show that our method produces more expressive maps than existing methods, with up to 10 times improvement in network reconstruction quality. We also show that our method extracts landmarks with more homogeneous nodes, with up to 90\% improvement in the average attribute/link entropy among the nodes over each landmark. Sense-making of a real-world network using a map computed by our method qualitatively verify the effectiveness of our method.
\end{abstract}

\section{Introduction}
\label{sec:intro}

Over the last decades, the size of datasets for real-world networks has expanded by orders of magnitude--- 
While this data deluge has blessed our building more robust analytics, it has challenged our ability to explore these large networks with intuition. 
Often, a network (e.g., Facebook) captures real-world information of entities as \emph{nodes} on the network, with \emph{attributes} (e.g., gender, age) describing the properties of and \emph{edges} indicating the  interactions (e.g., friend-link between users) between entities.
Such attributed, unweighted networks proliferate in many domains, e.g., social (e.g., Facebook, Twitter) and biology (e.g., protein-protein interaction) networks.
We thus address the problem of \emph{exploring large, real-world networks by creating compact representations that reveal the essential interactions in the data}, at desirable resolutions to see from the ``trees'' to the ``forest'' on the network.

We believe a compact representation for effectively exploring a network should satisfy the following requirements.
\emph{Conciseness}: The representation should preserve essential attribute and link patterns, while minimizing uninteresting detail, of the network. 
\emph{Expressiveness}: It should be able to encapsulate a diversity of characteristic patterns that may exist in the network.
\emph{Multi-resoultuons}: It can represent the network at various controllable levels of details.
 \emph{Locatability}: It should capture not only the big picture but also every node on the network-- i.e., we can locate each individual node on the representation.

\vspace{1.5mm}
\noindent{\bf Network cartography.\ }
The metaphor of a \textit{map} satisfies our requirements.
Maps, by definition of cartography~\cite{harley1987cartography}, are  {\it ``graphic representations that facilitate a spatial understanding of things, concepts, conditions, processes, or events in the human world.''}
In particular, for navigating the physical world, we are familiar with common geographic maps with ``landmarks'' and their spatial connections of ``roads.''
Such geographic maps are necessarily concise and expressive of all kinds of terrains, and allows zooming to various resolutions and locating every point of interest. 
Thus, for our objective, to compactly represent a large network, a \emph{network map} should comprise \emph{landmarks}, representing a group of individual nodes, and \emph{roads}, abstracting the interactions between these landmarks---
We aim to study this cartography of networks.

While the needs are pressing, unfortunately, there has been little prior work for network maps.
Rosvall et al.~\cite{rosvall2008maps} and Guimera et al.~\cite{guimera2005functional} made the early attempts to create maps to summarize complex networks. 
However, they either assume a specific domain (metabolic networks) or restrict to only link topologies. 
To our knowledge, our study is the first to create maps for general networks of both attribute and link information.
We informally state our problem as follows.

\vspace{2mm}
\noindent{\textbf{Problem:}}  For an attributed, unweighted network $G$, given the number of landmarks $K$, create a \emph{map} for $G$ as a graph $\set{M}_G$ with $K$ nodes to represent $G$.

\vspace{2mm}
\noindent{\textbf{Challenge:} \textit{Discovering landmarks}.}
What is a landmark on a network? 
While the important landmarks in the familiar geographical map have emerged due to centuries of cartographic practice, the semantics of the landmarks in a map designed to compactly represent a network are unclear.

\vspace{0.5mm}
\noindent{\textbf{Insight:} \textit{From homogeneity to roles.}\ } 
We observe that landmarks on a map signify \emph{homogeneity} of points of interests--- E.g., on a geographic map, a landmark (such as ``Navy Pier'') is an aggregate representation of points (shops, restaurants, piers) sharing similar properties (at very close spatial coordinates) and similar interconnections with other landmarks.
Thus, in a network, a landmark should aggregate nodes that are similar, in terms of their attributes and interactions.
Such homogeneous nodes, in \emph{sociology}, define \emph{roles} \cite{biddle2013role} in an interconnected society, as individuals that share characteristic patterns of behaviors. 
We thus propose to discover roles in a network as landmarks on its map.

\begin{figure}[!t]
	\centering
	\includegraphics[width=\linewidth]{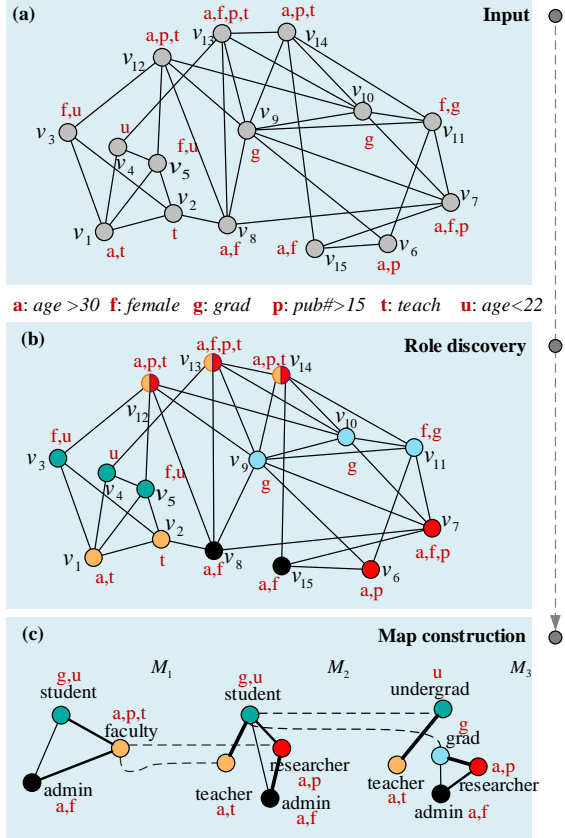}
	\vspace{-1cm}
	\caption{Example network $G_{ex}$ with node attributes (in red) representing the people and their interaction in an academic department. The middle figure colors each node by its role(s) resulted from role discovery. The bottom set of figures show a sequence of three maps with increasing resolution constructed from the discoverd roles, the relationships between the finer roles and those of the previous map are shown in dashed lines; each role is also labeled by the attributes it will likely have.}
	\vspace{-0.5cm}	
	\label{fig:ex}	
\end{figure}

To illustrate, Figure~\ref{fig:ex} shows an example network $G_{ex}$ in part (a) and the roles (as discovered via our algorithm) of the network in part (b).
In $G_{ex}$, each node is a person in an academic department, with attributes such as {\it a}:``age over 30'' and {\it t}:``teach'' for node $v_1$. 
While $G_{ex}$, a toy network of 15 nodes, already appears ``complex,'' its nodes share common patterns of behaviors.
To reveal, we annotate the network with the implicit roles in part (b) by coloring the nodes-- nodes with a color share a certain behavior pattern.
Now, we can observe that people in orange-role $R_3 = \{v_1,v_2\}$ ``teach and interact with green-role $R_4=\{v_3,v_4,v_5\}$.'' And, the latter group comprises young people.  
The red-role $R_1=\{v_6,v_7\}$ ``does not teach, has many publications, are aged over 30, and interacts with blue role $R_2=\{v_9,v_{10}, v_{11}\}$''.  
A plausible theory is that $R_4$ is \emph{undergrad} and $R_2$ \emph{grad} students;  then, $R_1$ and $R_3$ are perhaps \emph{researcher} and \emph{teacher}.
Further, with roles as landmarks, in part (c), we can construct maps of different resolutions, e.g., $\set{M}_1$ with three roles and $\set{M}_3$ five roles.

\vspace{2mm}
\noindent{\textbf{Challenge:} \textit{Formulating roles for landmarks}.} 
The challenge of formulating roles for landmarks lies in properly capturing the desired homogeneity. 
While there exist diverse formulations~\cite{everett1994regular,airoldi2009mixed,doreian2005generalized}, they fall short for landmark homogeneity, as Section \ref{sec:related} will further discuss.
First, they are too restricted for real-world networks, as none of them captures overlapping roles and inter-role interactions.
Second, they address only links but not attributes.
We thus need to develop formulation for coherently integrating the dual-mode behaviors in potentially overlapping roles.

\vspace{0.5mm}
\noindent{\textbf{Insight:} \textit{Probabilistic dual-mode regular equivalence.}\ } 
First, we conceptually capture link homogeneity by regular equivalence ({\sc Re}) \cite{borgatti1989class},  defined as {\it ''two nodes are regularly equivalent if they are equally related to equivalent others''}. This definition, however, is too restrictive for real-world networks. Our model for role extraction probabilistically captures the essence of {\sc Re} to handle noisy real-world data, and incorporates overlapping roles and non-uniform inter-role interactions to flexibly express the variations in complex networks.
Second, we inject attributes as the second mode of the role behavior, in addition to links. 
Thus, a role represents both link and attribute patterns, enforcing homogeneity in both modes.

\vspace{2mm}
\noindent{\textbf{Challenge:} \textit{Multi-resolution continuum}.} 
As maps should support multi-resolutions, maintaining an intuitive continuum between resolutions is challenging. 
On a geographic map, when we zoom into a region (e.g., Chicago), the changes are limited to \emph{within} the region  (we may now see Millennium Park), and the interconnections to the rest of the world should be stable (the whole region is still located in the state of Illinois).
Such continuum is intuitive for network maps too.
On map $\set{M}_1$ (Figure~\ref{fig:ex}c), when we zoom into landmark {\it faculty}, it gets refined into two finer landmarks in $\set{M}_2$: {\it teacher} and {\it researcher}, while the other landmarks are not changed; later, when we zoom into {\it student}, resulting in {\it undergrad} and {\it grad} in $\set{M}_3$, the continuum maintains.
While desired, realizing such continuum is challenging-- Since our landmarks are determined \emph{interdependently}, changing one landmark will necessarily change others.

\vspace{0.5mm}
\noindent{\textbf{Insight:} \textit{Multi-scale constraints.}}
To maintain continuum in multi-resolution views, as we use roles for landmarks, we must enforce the semantics of continuum in role discovery.
As we model role discovery as a likelihood maximization problem, we enforce the continuum as an additional constraint in the objective function, which thus enables a principled realization of multi-resolution stability.

\vspace{2mm}
\noindent{\textbf{Our solution.}
We summarize our framework of constructing a map for a given network $G$ as follows, as Figure~\ref{fig:ex} shows.
\vspace{2mm}
\mylistbegin	
\item {\bf Role discovery.} Compute $\theta$ as the {\it parameters of a model} that describes how the observed network $G$ is generated probabilistically from the latent roles (Figure~\ref{fig:ex}b).
\item {\bf  Map construction.} Use $\theta$ to compute and construct the map $\set{M}_G$ as a {\it node-labeled weighted graph}: each node of $\set{M}_G$ represent a role, the weight of the edge between two roles represent how likely they will interact, and each node is labeled with how likely that role will have each attribute (Figure~\ref{fig:ex}c).
\mylistend
\vspace{1mm}
Specifically, we propose a new probabilistic, generative model for networks to extract roles as landmarks of the map (Section~\ref{sec:model}). The model explicitly parameterizes each node-role affiliation pair and each role-role interaction pair.  The same role affiliation parameters will be the explanators in the logistic regression models for the node attributes. We design an objective function to optimize the likelihood of the observations under the model; our objective function formulation also encourages solution sparsity. Then, we process the model parameters into more intuitive quantities to construct the map, and achieve a multi-scale view by constraining the new solution to maintain the relationships and landmarks from the coarser scale, while determining the new landmarks at the finer scale and their relationship with the landmarks at the coarser scale (Section~\ref{sec:map}). We use an efficient block-coordinate gradient ascent algorithm to perform the optimization in linear time w.r.t the network size (Section~\ref{sec:algo}). 

In experiments, compared with baseline methods, ours shows significant improvement in model expressiveness measured by network reconstruction quality--- up to 10 times for synthetic networks and 27\% for real-world networks. In terms of homogeneity, our method outperforms existing methods by up to 90\% in terms the attribute/link entropy among the nodes over each landmark. In a case study using real-world networks, a set of multi-resolution maps created by our method actually help us to easily see the salient structures of the network from complex observed interactions.

\vspace{1.5mm}
\noindent{\textbf{Contributions.}  We summarize our contributions as follows.

\mylistbegin
    \item {\it Network cartography through role discovery.} We are the first to propose the use of roles to serve as map landmarks for attributed networks; in this, we are inspired by earlier work on regular equivalence~\cite{borgatti1989class}. The landmarks along with their attribute distribution and the resulting interrelationships between landmarks result in meaningful maps. 
    \item {\it A probabilistic model for the map.} We propose an expressive, probabilistic generative model of the network with roles as the latent factors. The expressive nature of the model allows us to model a wide variety of attributed networks.  During the optimization process, we enforce sparsity constraints to achieve concise maps.
    \item {\it Simultaneous multiple resolution.}  We are able to create maps that simultaneously show different parts of the map at different resolutions.
  	\item {\it Scalable cartography.} Our algorithm runs in linear time (with respect to network size). Furthermore, it is easy to implement using standard optimization toolboxes.
  	\item {\it Extensive experiments.} We not only synthesize networks with diverse structures to verify the effectiveness of our method in a wide range of networks, but also use real-world networks to show the practicality of our method, evaluated using a variety of measures. Case studies further evaluate our method's usability in action.
\mylistend

\section{Related Work}
\label{sec:related}
\noindent{\bf Network cartography.\ } Cartography~\cite{harley1987cartography} has a long history and was intensively used in visualizing geographic data, but their use in exploring large general networks has received attention only recently. Rosvall et al~\cite{rosvall2008maps} creates map of sciences by revealing community structures through efficiently encoding of random walk paths. Guimera et al ~\cite{guimera2005functional} used universal roles detected from optimizing network modularity to create maps for metabolic networks.

\vspace{1.5mm}
\noindent{\bf Attributed networks.\ }
We model node-attributed networks, for which alternative models with different goals have been proposed; e.g, {\sc Bagc}~\cite{xu2012model} uses a Bayesian approach, {\sc Cesna}~\cite{yang2013overlapping} models overlapping communities, and {\sc Mag}~\cite{kim2012latent} models clusters as generated from attributes. Methods for graph clustering using both structure/attribute information have also been proposed~\cite{zhou2009graph,silva2012mining}. Issues of these methods will be discussed later in this section.


\vspace{1.5mm}
\noindent{\bf Role theory.\ }
We create maps with landmarks as roles. The theory of roles have been studied in sociology~\cite{biddle2013role} where roles are defined as ``{\it a set of connected behaviors, rights, obligations, beliefs, and norms as conceptualized by people in a social situation}''~\cite{giddens2000introduction}, and various formulations have been proposed~\cite{burt1976positions,everett1994regular,borgatti1989class,mandel1983local}. For the alternative formulation of universal roles, which does not fit in our picture, see survey~\cite{rossi2015role}.



\vspace{1.5mm}
\noindent{\bf Intergroup interactions.\ }
Our model accounts for intergroup interactions in contrast with only intra-group found in communities. Some other methods also accounts for this aspect but with limitations for creating maps. Mixed-Membership Stochastic Blockmodeling ({\sc Mmb})~\cite{airoldi2009mixed} models each observed edge of each node as generated from one of its roles, but it does not model the intensity of role affiliation, as also pointed out in~\cite{yang2013overlapping,airoldi2009mixed}. ({\sc Bagc}~\cite{xu2012model} suffers from similar issues) as {\sc Mmb}. In fact, {\sc Mmb} cannot model hub and peripheral nodes, which in our model can be represented by nodes with strong affiliations to other roles. {\sc Mmb} does not model node attributes as key to information-enriched maps either. In terms of goals, {\sc Mmb} focus more on generalizability of model while we focus on summarization quality in computing maps. Generalized Blockmodeling~\cite{doreian2005generalized} does not capture varying interaction intensities between blocks (roles). {\sc Dedicom}~\cite{bader2007temporal} uses matrix factorization for real-valued data, thus cannot model binary links; other {\sc Nmf}-based methods~\cite{wang2011community,nickel2011three} suffers from similar issues. The model by Han et al~\cite{han2015probabilistic} uses separate parameters to explain node attributes and links.

\vspace{1.5mm}
\noindent{\bf Overlapping groups.\ }
Models for overlapping groups of nodes has less been studied. {\sc Dedicom}~\cite{bader2007temporal} and {\sc Cesna}~\cite{yang2013overlapping} are two examples. Note that mixed memberships~\cite{airoldi2009mixed,xu2012model} do not sufficiently capture the concept of overlapping groups, because their models only model the probability, rather than the intensity, of the affiliation of nodes to the groups.


\vspace{1.5mm}
\noindent{\bf Multi-resolution views.\ }
Local multiple-resolution views implies a hierarchical structure of the data (zooming corresponds to branching in a hierarchy). Methods for hierarchical clustering~\cite{friedman2001elements} cannot be directly applied to our case in which groups are overlapping. For graph summarization at different levels of granularity, Tian et al~\cite{tian2008efficient} proposed $k$-{\sc Snap}, but this operation does not support local views of part of the map.

\section{Role Discovery}
\label{sec:model}
In this section, we develop our probabilistic generative model of networks, with roles as latent factors, to extract the landmarks of a map.

\inlheader{Probabilistic regular equivalence}
We first identify {\sc Re} (regular equivalence) as the most appropriate semantics of link homogeneity for maps, compared to {\sc RolX}~\cite{henderson2012rolx} and structural equivalence~\cite{burt1976positions} ({\sc Se}) as representatives of alternative formulations. {\sc RolX} defines roles as {\it universal} in terms of structural functions, i.e, whether it is a bridge connecting otherwise separated groups, or a hub connecting to many nodes, etc.; it misses the interaction patterns of {\it specific} networks, thus falls short in expressness. {\sc Se} considers nodes equivalent (with same role) only if they connects to the same other {\it nodes} --- too restrictive since there can be too many nodes for {\sc Se} to find a small number of useful roles. On the other hand, {\sc Re} is free from both issues by considering nodes equivalent if they connects to the same roles rather than the same nodes.

We make significant extension of {\sc Re} with a probabilistic generative model, regarding roles as the latent factors that drives the noisy observed networks. For extracting latent factors~\cite{friedman2001elements}, although matrix factorizations (such as {\sc Nmf}~\cite{lin2007projected}) are also widely studied and used, they have not yet been well studied for effectively handling binary data matrices. Since binary data carry special semantics, thus methods for factorizing general real-valued data cannot be applied. From an efficiency perspective, {\sc Nmf} for a general $\mathbb{R}^{N\times N}$ data matrix (adjacency matrix in our case) needs $O(N^2)$ time~\cite{lin2007projected}, which is prohibitive for large networks; the {\sc Nmf}-based algorithm {\sc Asalsan}~\cite{bader2007temporal} specialized for graph processing has linear complexity but can only be used with weighted links. Probabilistic generative models, on the other hand, are a much more expressive class of tool which model various types of observations, including binary observations, e.g, by Bernoulli models. As shown later, an efficient algorithm can also be derived from a well designed probabilistic model.\\

\noindent{\bf Setting.\ } In this paper, we consider the extraction of $K$ landmarks given an undirected and unweighted graph $G=(\set{V}, \set{E}, \set{A})$ of $N$ nodes and $E$ links, where $\set{V}$ represents the node set and $\set{E}$ the edge set (directed graphs can be handled with slight modification of our method as described later; handling weighted graphs will be discussed till the end of the paper). Each node $v$ is labeled by $L$ binary attributes $a_{v,1}, \ldots, a_{v,L}\in \{0,1\}$, and we use $\set{A}$ to denote the set of observed node-attribute pairs $\{(v, i): a_{v,i}=1\}$. 

Regarding the notation, we will use boldface capital letters like $\matr{X}$ for matrices, calligraphic capital letters like $\set{X}$ for sets, boldface lowercase letters like $\vec{x}$ for vectors, and lowercase letters for scalars. The roles will be indexed by $k$ and the attributes by $i$.

\subsection{Model Description}
We model the landmarks in a map as roles. In the example network $G_{ex}$ in Figure~\ref{fig:ex}, we observe group $g_1=\{v_6, v_7\}$ ``often interact with ($g_2=\{v_9, v_{10}, v_{11}\}$)'' and ``have more than 15 publications'' but ``do not teach courses''. A plausible theory would be that $g_1$ are research professors and $g_2$ graduates, and $g_1$'s behaviors can be explained by the {\it pure} role of ``researcher''. We also observe ``aged over 30'' and ``frequent interactions with group $g_3=\{v_3,v_4,v_5\}$'' for group $g_4=\{v_1, v_2\}$; prior knowledge tells us $g_4$ are lecturers and $g_3$ are undergrads, and $g_4$'s behaviors are due to the pure role of ``teacher''. Other roles include ``admin'', ``undergrad'', and ``graduate'', as extracted in map $\set{M}_3$ in Figure~\ref{fig:ex}.

\inlheader{Plural and non-uniform role affiliation}
To characterize the roles, we observe that each node can participate in {\it multiple} roles at different intensities. First, role affiliation is {\it plural}. In $G_{ex}$, group $g_5=\{v_{12},v_{13}, v_{14}\}$ (known as professors in prior) do both teaching and research and interact with both undergrad and graduate students; these behaviors are explained by their simultaneous roles ``teacher'' (abbrv. {\it tea}) and ``researcher'' (abbrv. {\it res}). Second, the affiliation is {\it non-uniform}. Note that professor $v_{12}$ might not conduct role ``teacher'' so much as lecturers as indicated by less interactions with undergrads.

To capture the plurality and non-uniformity of role affiliation, we represent the affiliation of each node $v$ to a set of $K$ roles by vector $\vec{x}_v=(x_{v,1}, \ldots, x_{v,K})$, where $x_{v,k}$ quantifies how strongly $v$ is affiliated to the role $k$; we also require $\vec{x}_v\geq 0$ for easier interpretation of the results. For instance, in $G_{ex}$, the lecturer $v_1$ will have dominantly large $x_{1,tea}$ compared to other components of $\vec{x}_1$ due to a pure role of ``teacher'', while the professor $v_{12}$ who does research but less teaching will have $x_{12, res}>x_{12, tea}> 0$. Stacking $\vec{x}_v^T$ form the {\it node-role membership matrix} $\matr{X}\in\mathbb{R}_{\geq 0}^{N\times K}$.

\inlheader{Non-uniform inter-role interaction}
We observe that inter-role interactions are also {\it non-uniform} in intensities. In $G_{ex}$, although $g_6=\{v_8, v_{15}\}$ (married female aged over 30, possibly administrative staff) interact with both $g_5$ (professors) and $g_2$ (graduates), administrative staff usually interact more with professors, because they often assist the professors in scheduling events.

To reflect the non-uniformity of inter-role interactions, we quantify the tendency for two nodes of roles $k$ and $l$ to interact by $r_{k,l}\geq 0$. We then define the {\it inter-role interaction matrix} as $\matr{R}=(r_{i,j})\in\mathbb{R}_{\geq 0}^{K\times K}$.  For instance, in $G_{ex}$, the strong interaction between professors and graduates (abbrv. {\it gra}) will be captured by $r_{res,gra},r_{gra,res}>0$ and possibly also $r_{gra,res}>r_{res,gra}$ as graduates usually more actively turn to professors.


Based on the above notions, we model our belief for two nodes $u,v$ interact as follows. We know that if nodes $u, v$ are affiliated to roles $k$ and $l$ respectively and roles $k$ and $l$ tends to interact, nodes $u$ and $v$ also tends to interact. In addition, our belief for $(u,v)$ being formed will add up as we observe more such evidences from other pairs of roles $k',l'$. For simplicity, we assume that our belief grows linearly with the role affiliation and inter-role interaction intensities. Thus we formulate this belief, the {\it predictor} of link $(u,v)$, as
$$
\rho_{u,v}\coloneqq \sum_{k,l\in [K]} x_{u,k} \cdot r_{k,l} \cdot x_{v,l} = \vec{x}_u^T \matr{R} \vec{x}_v 
$$
Since our observations are binary, we need a {\it link function} that transform $\rho_{u,v}$ into probability values. Observing that the marginal potential of forming links normally decrease as more evidences are present, we model the probability of forming link $(u,v)$ with the predictor $\rho_{u,v}$ as
\begin{equation}
\label{eq:plink}
p_{u,v} \coloneqq \varphi(\rho_{u,v}) = 1-\exp(-\vec{x}_u^T \matr{R} \vec{x}_v)
\end{equation}
While another common choice for transforming predictors to have range $(0,1)$ would be the sigmoid function $\sigma(\cdot)$, it has domain $(-\infty, \infty)$ but $\rho_{u,v}$ is restricted to be nonnegative, hence inappropriate for our case. We also support our use of $\varphi(\cdot)$ by efficiency considerations as we will discuss in Section~\ref{sec:algo}.

\inlheader{Links and attributes as covariates}
Finally, we observe that the links and attributes of each node are {\it covariates} with its roles as the underlying factor. In $G_{ex}$, the attribute ``age > 30'' and interactions of ``being approached by graduate students'' of the professor $v_12$ can both be explained by a ``researcher'' role. Similarly, the attribute ``female'' and the interactions of ``being contacted by professors'' can be explained by a ``admin'' role.

To capture the role-based covariation of attributes and links, we will use an attribute predictor as the linear combination of the role affiliation vector $\vec{x}_v$ and some parameter $\vec{w}_i$ to model our belief of node $v$ to have attribute $i$ (for simplicity, we assume that the attributes of a node are conditionally independent given the roles of that node). For example, in $G_{ex}$, the role ``researcher'' will indicate a high weight $w_{res,(pub>15)}$ so that it would contributes positively and strongly to observing many publications for professors. Then, we transform the predictor to probability using a link function. For the above purpose, a standard logistic regression would suffice, by which we specify the attribute predictor as $\mu_{v,i}\coloneqq \vec{w}_i^T \vec{x}_v$ and the probability of observing $a_{v,i}=1$ as
\begin{equation}
\label{eq:pattrib}
q_{v,i} \coloneqq \sigma(\mu_{v,i}) = \frac{1}{1+\exp(-\vec{w}_i^T \vec{x}_v)}
\end{equation}
Note the intercept term is ignored in the predictor to simplify our presentation. We denote $\matr{W}=(\vec{w}_1,\ldots,\vec{w}_L)$.\\



Now, our problem of discovery roles as landmarks of the map for a given network is defined as follows.
\begin{defn}[Role discovery]
	Given an undirected unweighted graph $G$ with node attributes and the number of roles $K$, we will find a set of roles represented in the following parameters: the matrix $\matr{R}$ for inter-role interactions, a vector $\vec{w}_i$ for each attribute $i$ for predicting the attributes of any node $v$ given the roles of $v$, and a vector $\vec{x}_v$ for each node $v$ for representing the role affiliations of $v$.
\end{defn}

\noindent{\bf Model extensibility.\ }
Although binary attributes are discussed here, the model can be easily adapted to categorical attributes (e.g, occupation) using a multinomial logistic model in place of the binary model described earlier. The model also directly applies to directed graphs which will result in an asymmetric matrix $\matr{R}$.
To support weighted graphs, we can modify the model by replacing $\varphi(x)$ in \eqref{eq:plink} with an identity function, while the algorithm needs to be designed differently (e.g, the {\sc Asalsan} algorithm~\cite{bader2007temporal} with some adaption), for which we will not discuss in more details here.



\subsection{Objective Function}
We extract the roles from network $G$ by optimizing the parameters of our model of $G$ in the following aspects. First, we optimize the parameters such that they {\it accurately express} the observations in the original network $G$ using a set of roles. Since the number of landmarks $K$ is usually small for concise maps, it implies that each landmark must be homogeneous for the observations to be represented compactly using the roles. In particular, we capture the accuracy of $G$ by how well our model can be used to reconstruct $G$. Second, we optimize for {\it sparsity} of the parameters for conciseness of representation. That is, we wish $G$ be represented using as few inter-role interactions as possible and each node can be described by their most important roles, such that users can focus on the most salient patterns of $G$.

To encode the above requirements, we propose an objective function $f(\theta)$ with respect to parameters $\theta=\{\matr{X}, \matr{R}, \matr{W}\}$. $\theta$ captures model accuracy by the log-likelihood of observing $G$ under a model with parameters $\theta$, while the sparsity requirement is enforced by the $\ell_1$-norm regularizations of $\matr{R}$ and $\matr{X}$. Formally,
\begin{align}
f(\theta) = (1-\alpha)\ell_{\set{E}} + \alpha \ell_{\set{A}}
-\alpha_R \|\matr{R}\|_1 - \alpha_X \|\matr{X}\|_1 \label{eq:obj}
\end{align}
where $\ell_{\set{E}}$ and $\ell_{\set{A}}$ are the log-likelihoods of links and attributes given by
\begin{align}
\ell_{\set{E}} &= \sum_{(u,v)\in \set{E}} \log\varphi(\rho_{u,v}) + \sum_{(u,v)\notin \set{E}} \log(1-\varphi(\rho_{u,v})) \label{eq:llink} \\
\ell_{\set{A}} &= \sum_{(v,i)\in \set{A}} \log\sigma(\mu_{v,i}) +
\sum_{(v,i)\notin \set{A}} \log(1-\sigma(\mu_{v,i})) \label{eq:lattrib}
\end{align}
and weight $\alpha$ is used to impose emphasis on links or attributes, and $\alpha_X, \alpha_R$ to control the degree of representation sparsity.

In summary, the optimal parameters $\theta^*$, which maximizes $f(\theta)$ subject to the constraints that $\matr{X}$ and $\matr{R}$, will represent our extracted roles. That is,
\begin{align}
\label{eq:problem}
\theta^* = \argmax_{\matr{X}\geq 0,\matr{R}\geq 0, \matr{W}} f(\theta)    
\end{align}
The detailed algorithm for optimizing $f(\theta)$ will be introduced later in Section~\ref{sec:algo}.

\section{Map Construction}
\label{sec:map}
Thus far we have encoded the information needed to create a map for the given network in the model parameters $\theta$. In this section, we will develop methods to render $\theta$ as an intuitively interpretable map. Further, we propose methods to allow for local multiple-resolution views of any part of the map.
%

\subsection{From Model to Map}
The computed model parameters $\theta^*$ in the stage of role discovery needs to be transformed to interpretable quantities to construct a map. For instance, while a larger $x_{v,k}$ apparently indicates the stronger affiliation of node $v$ to role $k$, $x_{v,k}$ does not have an intuitive physical meaning, such as probability, length, or frequency which are natural for human interpretation. The same issue also holds for parameters $\vec{x}_v$ and $\vec{w}_i$. This makes it non-intuitive, if not deceptive, to make sense of a map constructed directly from $\theta^*$ (e.g, using $\matr{R}$ as the weighted adjacency matrix for the metagraph).


In order to construct maps as metagraphs with interpretable weights and labels, we introduce a {\it virtual node} $v_k$ for each role $k$, which is an {\it imaginary} node affiliated purely to role $k$, i.e., $\vec{x}_{v_k}=c\hat{\vec{e}}_k$ for some $c>0$, where $\hat{\vec{e}}_k$ is the unit vector along the $k$th dimension, and $c$ controls the affiliation strength of $v_k$ to role $k$. Intuitively, the virtual node $v_k$ can be viewed as an idealized representative for role $k$ so that its behaviors as expected from our model naturally characterize role $k$. We might choose the value of parameter $c$ as follows. First, we may set $c$ to some constant independent of the actual affiliations of the nodes to role $k$, thus in effect characterize role $k$ {\it itself} by $v_k$. Second, we may set $c=\bar{x}_{\cdot, k}=(\sum_{v\in \set{L}(k)} x_{v,_k})/|\set{L}(k)|$, where $\set{L}(k)=|\{v: x_{v,k}>0\}|$, as the average affiliation of nodes with role $k$; such created virtual node $v_k$ will instead represent the role $k$ as actually practiced by the real nodes in data.

Using the virtual node $v_k$ as a representative for role $k$, we characterize the behaviors of role $k$ intuitively by the probability of observing each attribute $i$ on $v_k$, given by
$$
\psi_{k,i}\coloneqq \sigma(\vec{w}_i^T \hat{c\vec{e}}_k)=(1+\exp(-c\cdot w_{k,i}))^{-1}
$$
Similarly, to characterize the inte-role interaction between each pair of roles $k$ and $l$, we use the probability of interaction between an virtual node $v_k$ with purely role $k$ and a virtual node $v_l$ with purely role $l$, which we give by
$$
\omega_{k,l}\coloneqq \varphi(c\hat{\vec{e}}_k^T\matr{R} c\hat{\vec{e}}_l)=1-\exp(-c^2 r_{k,l})
$$
For later reference, we denote $\matr{P}_{attrib}=(\psi_{k,i})_{(k,i)\in [K]\times [L]}$ and $\matr{P}_{link}=(\omega_{k,l})_{(k,l)\in [K]^2}$. Such transformed $\theta$ will be finally visualized as the output map.

In summary, we formally define the steps for creating a map as follows.
\begin{defn}[Map construction]
Given an unweighted network $G$ with $L$ binary node attributes, a map for $G$ with $K$ landmarks as a metagraph $\set{M}_G$ is constructed by
\begin{enumerate}
  \setlength{\itemsep}{-1pt}
  \item Compute $\theta^* =\argmax f(\theta)$ given $G$;
  \item Use $\theta^*$ to compute $\psi_{k,i}$ for each $(i,k)\in [L]\times [K]$ and $\omega_{k,l}$ for each $(k,l)\in [K]^2$;
  \item Construct graph $\set{M}_G=(\set{R}, \set{I}; \psi_{\cdot,\cdot}, \omega_{\cdot,\cdot})$: each node in $\set{R}$ represents a landmark, each edge $(k,l)\in\set{I}$ is weighted by $\omega_{k,l}$, and each node $k\in\set{R}$ is labeled with $\psi_{k,i}$ for each attribute $i$.
\end{enumerate}
\end{defn}


\subsection{Local Multi-Resolution Views}
With an overview of the network by a coarser map, we allow users to select {\it part} of the map for higher-resolution views in the same global context, as in map $\set{M}_2$ in Figure~\ref{fig:ex} where we selected the landmark ``student'' to zoom in, and revealed the finer-granular landmarks ``undergrad'' and ``graduate'' in $\set{M}_3$, while other landmarks remains unchanged, and the finer-granular connections between the new landmarks and the existing landmarks are revealed too, such as that ``graduate'' interacts more with professors than ``undergrad''.

As a first attempt to changing the resolution locally, one might extract a subgraph for the selected landmark and compute a map with that subgraph as input, but this method loses the connections between the newly created landmarks with those existing outside the subgraph. One might also attempt the alternative way of computing from scratch a new map with more landmarks, but it violates our locality requirement and destructs the existing non-selected landmarks. Although hierarchical clustering also produces finer-granular views of data, it does hard partitioning of the data points while our groups of nodes are overlapping; it also considers only intra-cluster connections but not the connections between the clusters.

\begin{figure}[ht]
\centering
\includegraphics[width=0.7\linewidth]{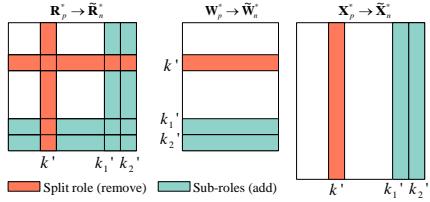}
\vspace{-3mm}
\caption{Obtaning $\tilde{\theta}_p^*$ from $\theta_p^*$ to initialize $\theta_n^*$}
\label{fig:reduce}
\end{figure}

We achieve local multiple-resolution views for some role $k'$ in a given map $\set{M}_p$ constructed with $\theta_n^*$ by constructing a new map $\set{M}_n$ with parameters $\theta_n*$, and constraining our optimization of $\theta_n^*$ such that the non-selected roles (those other than $k'$) in $\set{M}_p$ remain little changed. We measure the change between maps $\set{M}_p$ and $\set{M}_n$ by the distance between their respective parameters $\theta_p^*$ and $\theta_n^*$ with respect to the non-split roles. Formally, we define the distance between models $\theta_n$ and $\theta_p$ using sum of squared errors as
$$
d_{map}(\theta_p, \theta_n) = \|\reduce_{k'_1,k'_2}(\theta_n)- \reduce_{k'}(\theta_p)\|_2^2
$$
where $\reduce_{k'}(\theta)$ (resp. $\reduce_{k'_1,k'_2}(\theta)$) transforms $\theta$ by removing its components associated with role $k'$ (resp. $k'_1, k'_2$), for which we illustrate the details in Figure~\ref{fig:reduce}. Using the operator $\reduce_{k'}(\cdot)$, we actually measure the divergence between the maps only with respect to roles other than the selected $k'$.

In summary, we formally define the problem of computing local multi-resolution views on a map as follows.
\begin{defn}[Local multi-resolution views]
Given a previously computed map $\set{M}_p$ constructed from $\theta_p^*$ with $K$ roles, and a specified landmark $k'$ to view at finer resolution, we compute a map $\set{M}_p$ with model $\theta_n^*$ of $K+1$ roles by splitting role $k'$ into the {\it sub-roles} $k'_1$ and $k'_2$, such that $\theta_n^*$ satisfies our requirements for appropriate landmarks by maximizing $f(\theta_n^*)$, meanwhile keeps the roles other than $k'$ relatively unchanged by minimizing $d_{map}(\theta_p^*, \theta_n^*)$
\end{defn}
Note that while we demonstrate the case of two sub-roles, the algorithm can be easily generalized to split any role $k'$ into multiple sub-roles.


In order to find $\theta_n^*$ that both maximizes $f(\theta_n^*)$ for quality landmarks and minimizes $d_{map}(\theta_p^*, \theta_n^*)$ for locality of the finer view, we optimize $\theta_n^*$ under a revised objective function
\begin{equation}
\label{eq:obj-zoom}
f_z(\theta_n; \theta_p^*)\coloneqq f(\theta_n) - \beta\cdot d_{map}(\theta_p^*, \theta_n)
\end{equation}
where $\beta$ is the hyperperameter controlling the tradeoff between the landmark quality and the locality of the resolution change. For the revised optimization objective, we will design an efficient algorithm in Section~\ref{sec:algo}.


\section{Algorithms}
\label{sec:algo}
In this section, we develop efficient algorithms to optimize the objective functions formulated in Section~\ref{sec:model} for role discovery and Section~\ref{sec:map} for computing local multi-resolution views in selected landmarks.\\

\noindent{\bf Algorithm for landmark extraction.\ }
Given the optimization objective function $f(\theta)$ formulated in \eqref{eq:obj}, we derive the gradients of $f(\theta)$ with respect to variables $\matr{X},\matr{R},\matr{W}$ as follows. For easier derivation, we split the matrix variables $\matr{X}$ and $\matr{W}$ into the vector variables $\vec{x}_v$ and $\vec{w}_b$. Then
\begin{align}
\frac{\partial f}{\partial \matr{R}} &=
(1-\alpha)\bigg(\underbrace{\sum_{(u,v)\in \set{E}} \frac{1-\varphi(\rho_{u,v})}{\varphi(\rho_{u,v})} \vec{x}_u \vec{x}_v^T}_{\text{present links}} - 
\underbrace{\sum_{(u',v')\notin \set{E}} \vec{x}_{u'} \vec{x}_{v'}^T}_{\text{absent links}} \bigg) \nonumber\\
& - \alpha_R\cdot \sign(\matr{R}) \label{eq:gr} \\
\frac{\partial f}{\partial \vec{x}_v} &= (1-\alpha)\bigg(\underbrace{\sum_{(v,u)\in\set{E}} \frac{1-\varphi(\rho_{u,v})}{\varphi(\rho_{u,v})} \matr{R} \vec{x}_u}_{\text{neighbors of } v} -
\underbrace{\sum_{(v,u')\notin\set{E}} \matr{R} \vec{x}_{u'}}_{\text{non-neighbors of } v} \bigg) \nonumber \\
& + \alpha \bigg(\sum_{(v,i)\in\set{A}} (1 - \sigma(\mu_{v,i}))\vec{w}_i -
\sum_{(v,i')\notin\set{A}} \sigma(\mu_{v,i'})\vec{w}_{i'} \bigg) \nonumber \\
& - \alpha_X \sign(\vec{x}_v) \label{eq:gx} \\
\frac{\partial f}{\partial \vec{w}_i} &=
\alpha \bigg(\sum_{(v,i)\in \set{A}} (1 - \sigma(\mu_{v,i})) \vec{x}_v -
\sum_{(v,i')\notin \set{A}} \sigma(\mu_{v,i'})\vec{x}_v \bigg) \label{eq:gw}
\end{align}
Note that the optimization problem \eqref{eq:problem} is {\it nonconvex}. The problem of optimizing the variable $\matr{R}$ with all other variable fixed, however, is convex. This is also true for variables $\vec{x}_v$ and $\vec{w}_i$. Thus we resort to the {\it block-coordinate gradient ascent} algorithm to optimize $f(\theta)$, where we cycle through the variables and optimize one of them at a time until convergence to a local minima. To optimize each variable, we employ the gradient ascent algorithm with the learning rate set proportional to the normalized gradient. After each gradient ascent update of any variable $z$ subject to the non-negativity constraint, we need an additional step to ensure the updated $z'$ still satisfies the constraint. To this end, we project the negative entries of $z'$ to $0$ after each iteration, denoted by $z'\leftarrow \proj^+(z)$. The steps are summarized in Algorithm~\ref{alg:netcart}.\\

\begin{algorithm}[!t]
	\label{alg:netcart}
	\SetKwInOut{Input}{Input}
	\SetKwInOut{Output}{Output}
	\Input{Network $G$, number of landmarks $K$}
	\Output{Landmarks $\theta^*=(\matr{X}^*$, $\matr{R}^*$, $\matr{W}^*)$}
	
	\BlankLine
	\tcp{initialization}
	Initialize entries of $\matr{X},\matr{W}$ with small random values\;
	Initialize $\matr{R}$ with $(\matr{M}+\matr{M}^T)/2$ for some matrix $\matr{M}$ with small random entries\;
	\Repeat{$f$ not improved}{
		\tcp{evaluate gradients with precomputed $\tilde{\vec{x}}$}	
		$\tilde{\vec{x}}\leftarrow \sum_v \vec{x}_v$\;
		Repeat $\matr{R}^{(t+1)} \leftarrow \proj^+(\matr{R}^{(t)} + \beta^{(t+1)}\frac{\partial f}{\partial \matr{R}^{(t)}})$\;
		\tcp{enforcing symmetry of $\matr{R}$ by projection}
		\If{$G$ is undirected}{
			$\matr{R}\leftarrow (\matr{R}+\matr{R}^T)/2$\;
		}
		\For{each node $v$}{
			Repeat $\vec{x}_v^{(t+1)} \leftarrow \proj^+(\vec{x}_v^{(t)} + \beta^{(t+1)}\frac{\partial f}{\partial \vec{x}_v^{(t)}})$\;
			Update $\tilde{\vec{x}}$ incrementally due to new $\vec{x}_v$\;
		}
		\For{each attribute $i$}{
			Repeat $\vec{w}_i^{(t+1)} \leftarrow \vec{w}_i^{(t)} + \beta^{(t+1)}\frac{\partial f}{\partial \vec{w}_i^{(t)}}$\;
		}        
	}
	\caption{\algrole}
\end{algorithm}

\noindent{\bf On Initialization.\ }
At the beginning of the optimization, we need to initialize each variable properly. Initializing the variables to small uniformly distributed random values around $0.1$ turns out to work well in practice. An initialization with large random values, however, will typically lead to useless results with the algorithm stopping at an early stage. In addition, special attention must be paid to the $\matr{R}$ when the given network $G$ is undirected. In this case, $\matr{R}$ needs to be initialized as some {\it symmetric} matrix. It can be easily shown that the symmetry property will then continue to hold throughout the optimization. In practice, setting $\matr{R}$ to $(\matr{M}+\matr{M}^T)/2$ for some random matrix $\matr{M}$ will do the work, while failures in confirming to symmetry will result in unexpected nonsense results in most cases.\\


\noindent{\bf Linear-time computation.\ }
The cost of Algorithm~\ref{alg:netcart} concentrates on the evaluation of the log-likelihoods \eqref{eq:llink}, \eqref{eq:lattrib} for evaluating the objective function and the gradients \eqref{eq:gr}--\eqref{eq:gw}. To analyze the time for objective function evaluation, it amounts to analyzing the evaluation of  $\rho_{u,v}$ for each link $(u,v)\in\set{E}$ individually and the sum of $\rho_{u',v'}$ for all absent links $(u',v')\notin\set{E}$.

Simply evaluating $\rho_{u,v}$ for each pair of vertices is prohibitive. While for each observed link $(u,v)$, the term $\rho_{u,v}$ takes $O(K^2)$ time for evaluating $\vec{x}_u^T \matr{R} \vec{x}_v$, there are $O(N^2)$ separated pairs of nodes, for which computing $\rho_{u',v'}$ once for each such pair is impossible.

Fortunately, the sum of $\rho_{u',v'}$ for missing links can be improved by first rewriting it into 
$$
\sum_{(u',v')\notin \set{E}} \rho_{u',v'} = \sum_{(u,v)\in \set{V}^2} \vec{x}_{u}^T \matr{R} \vec{x}_{v} - \sum_{(u,v)\in \set{E}} \rho_{u,v}
$$
We observe that the first term on the right side can be simplified into $\tilde{\vec{x}}^T \matr{R} \tilde{\vec{x}}$ where $\tilde{\vec{x}} =\sum_{v\in\set{V}} \vec{x}_v$. The value of $\tilde{\vec{x}}^T \matr{R} \tilde{\vec{x}}$ can be precomputed in $O(NK+K^2)$ time and cached before optimizing $\matr{R}$ in each outer iteration. The total time for evaluating $f(\theta)$ is then $O((N+E)K^2)$. The costly last term in \eqref{eq:gr} can be similarly computed.

With the precomputation technique, the time for a full (outer) iteration of the algorithm scales {\it linearly} w.r.t the size of $G$. In details, each gradient ascent step for $\matr{R}$ takes $O((N+E)K^2)$ time, each $\vec{x}_v$ takes $O(d_v K^2 + LK)$ time where $d_v$ is the degree of node $v$, and each $\vec{w}_i$ takes $O(NLK)$ time. Summing up, the overall time complexity of Algorithm~\ref{alg:netcart} is $ O((N+E)K^2 + NLK)$; the memory needed is $O(NK+K^2+KL)$ proportional to the size of $\theta$. The performance in practice can be further boosted using optimized matrix operation libraries.

Note that the precomputation technique is valid for our choice of $\varphi(x)$ as the link function but {\it not} for the logistic function $\sigma(x)$. With $\sigma(x)$, term $\rho_{u,v}$ has to be evaluated {\it once for each} such link. In fact, with $\sigma(x)$, term $\sum_{(u',v')\notin\set{E}} \vec{x}_{u'} \vec{x}_{v'}^T$ in \eqref{eq:gr} will be replaced by $ \sum_{(u',v')\notin\set{E}} \sigma(\rho_{u',v'})\cdot \vec{x}_{u'}\vec{x}_{v'}^T$ which cannot be decomposed as the product of two cheap sums as in case of $\varphi(x)$ mean function. This results in prohibitive $O(N^2)$ complexity.\\

\noindent{\bf Algorithm for zooming.\ }
We create locally refined view $\theta_n^*$ with respect to the selected landmark $k'$ in a map with parameters $\theta_p^*$ by optimizing $f_z(\theta_n; \theta_p^*)$ as formulated in \eqref{eq:obj-zoom}. Let the sub-roles to be created be $k_1'$ and $k_2'$. We derive the gradients of $f_z(\theta_n; \theta_p^*)$ as follows.
$$
\frac{\partial f_z}{\partial \theta_n} = \frac{\partial f}{\partial \theta_n} - 2\beta\cdot\mask_{k_1',k_2'}(\theta_n - \tilde{\theta}_p^*)
$$
where $\tilde{\theta}_p^*$ is the extended $\theta_p^*$ which will be used to initialize $\theta_n$. In particular, we create $\tilde{\theta}_p^*$ from $\theta_p*$ by removing its parameters for the role $k'$ and appending new (randomly initialized) parameters for the sub-roles $k'_1$ and $k'_2$; see  Figure~\ref{fig:reduce} for an illustration of this process. The operator $\mask_{k_1',k_2'}(\theta_n)$ sets the parameters in $\theta_n$ corresponding to the sub-roles $k'_1,k'_2$ to 0. Intuitively, $\mask(\cdot)$ keeps the newly created sub-roles from being penalized by its distance from $\tilde{\theta}_p^*$, . The weight $\beta$ controls the degree of the closeness enforcement. 

We optimize $\theta_n^*$ using Algorithm~\ref{alg:netcart} with the adapted gradients and also a special initialization of $\theta_n$ for {\it faster convergence}. In particular, we initialize $\theta_n$ to $\tilde{\theta}_p^*$ which inherits the roles other than $k'$, so that the algorithm's overhead will be limited to the optimization of the sub-roles. The algorithm for zooming is summarized in Algorithm~\ref{alg:zoom}.

\begin{algorithm}[!t]
	\label{alg:zoom}
	\SetKwInOut{Input}{Input}
	\SetKwInOut{Output}{Output}
	\Input{Network $G$, landmarks $\theta_p^*$, selected landmark $k'$}
	\Output{Landmarks $\theta_n^*$ with refined $k'$}
	\caption{Zoom}
	
	\BlankLine
	$\tilde{\theta}_p^*\leftarrow \reduce_{k'}(\theta_p^*)$\;
	Append parameters for $k'_1, k'_2$ to $\tilde{\theta}_p^*$ and initialize them with small random numbers\;
	Initialize $\theta_n$ by $\tilde{\theta}_p^*$ \;
	Compute $\theta_n^*= \argmax_{\theta_n} f_z(\theta_n; \theta_p^*)$ using Algorithm~\ref{alg:netcart} with adapted gradients
\end{algorithm}

%


\section{Experimental Evaluation}
\label{sec:eval}
We will both quantitatively verify if the maps computed by our method have the desired properties of maps we identified in Section~\ref{sec:intro}. We will also demonstrate the usability of our method by creating multi-resolution maps for real-world networks.

\inlheader{Datasets}
The following real-world and synthetic datasets will be used in our tests. We chose the real networks so that they exhibit interesting strong intergroup interactions instead of only intra-group interactions found in communities.
\vspace{-2mm}
\begin{citem}
	\item {\tt Hospital}~\footnote{\small \url{http://www.sociopatterns.org/datasets}}. This dataset~\cite{vanhems2013estimating} contains the directed network of contacts among 75 people in a hospital ward in France during five days. Each person is labeled with one of \{{\tt patient, admin, nurse, doctor}\}. An arc is created for each pair of people if they interacted at least 5 times during the period.
	\item {\tt Enron}~\footnote{\small \url{http://cis.jhu.edu/~parky/Enron}}. This dataset represents the directed network of email communications between 184 users. Each user is labeled by one or more the 32 topics (manually extracted in \cite{berry2001topic}) they ever used. We create an arc for each (ordered) pair of users if they communicated at least 15 emails.
	\item {\tt Syn-x}. We synthesize $\theta$ as parameters of our proposed model with $K=5$ landmarks, which in turn generates random undirected networks, such that the network corresponds to one of \{{\tt bip,star,comm,rand}\} representing bipartite, star (wheel), community, and random structures of $G_{map}$ (constructed from $\matr{R}$). Thus, a variety of common graph structures will be captured in the experiments. We then synthesize networks according our probabilistic generative model with parameter $\theta$.
\end{citem}
\vspace{-2mm}

\inlheader{ Methods}
We will evaluate and compare the following methods.
\vspace{-2mm}
\begin{citem}
	\item \algrole. Our model of networks for creating maps. By default, we set the weight for attribute likelihoods as$\alpha=0.5$ and sparsity regularization weights $\alpha_R,\alpha_X$ to 0 when compared with other method and 0.2 otherwise.
	\item {\sc Cesna}~\cite{yang2013community}. This the state-of-the-art model for large node-attributed networks. It aims at overlapping groups, models attributes as generated from group memberships, as well allows flexible community affiliation. By default, $\alpha=0.5$ which has the same role as ours. In {\sc Cesna}, $K$ represents the number of communities, analogous to our landmarks.
	\item {\sc BinDedi}. The original {\sc Dedicom}~\cite{bader2007temporal,harshman1982model} model uses a similar form $\matr{A}\approx \matr{X}\matr{R}\matr{X}^T$ to model links, which allows arbitrary membership affiliation and pairwise intergroup interactions, but handles only weighted graphs and does not handle attributes. We non-trivially adapt it for binary links by applying the link function $\varphi(\cdot)$ to $\matr{X}\matr{R}\matr{X}^T$ and $\matr{A}\sim Bernoulli(\varphi(\matr{X}\matr{R}\matr{X}^T))$. The modified {\sc Dedicom} is equivalent to our model with $\alpha=0$.	
\end{citem}


\subsection{Comparing with Baselines}

\inlheader{Convergence}
We evaluate the convergence performance of Algorithm~\ref{alg:netcart} as indicator for efficiency. Figure~\ref{fig:exp-conv} shows how the value of the objective function $f(\theta)$ changes with the number of full iterations. We observe that the algorithm improves significantly at the first few iterations and converges quickly. Thus our algorithm is scalable given also the linear time complexity of each iteration.
\begin{figure}[!t]
	\centering
	\includegraphics[width=\linewidth]{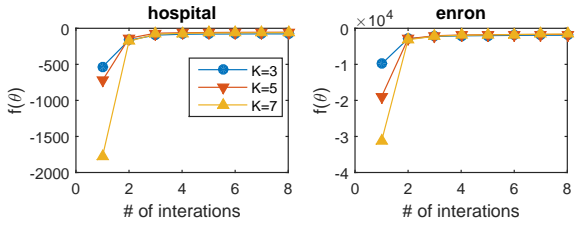}
	\vspace{-0.5cm}
	\caption{Convergence of \algrole (Algorithm~\ref{alg:netcart}).}
	\vspace{-0.5cm}	
	\label{fig:exp-conv}
\end{figure}

\inlheader{Expressiveness}
We measure a map's expressiveness by quality of network reconstruction using the map $\theta^*$, in terms of the log-likelihood of observing the network given the model $\theta^*$ underlying our map. In Figure~\ref{fig:exp-ll-real}, we show the log-likelihoods of the real-world networks at different $K$ computed by \algrole and {\sc Cesna} (method {\sc BinDedi} is not considered here due to its inability to handle node attributes; it corresponds to \algrole with $\alpha=0$ and the effect of $\alpha$ will be evaluated later). We observe that our method outperforms {\sc Cesna} by up to $\sim 27\%$ on {\tt hospital} and $\sim 13\%$ on {\tt enron} in reconstruction quality. For both methods, the log-likelihoods improve with larger $K$, as expected due to more parameters.

To evaluate the reconstruction quality of \algrole on networks with diverse structural characteristics, we further use four synthetic networks with structures often found in real-world networks, as shown in Figure~\ref{fig:exp-ll-syn}. For {\tt syn-star}, we observe that \algrole drastically outperforms {\sc Censa} by orders of magnitude. Obvious advantage of our method is also observed for {\tt syn-bip} and {\tt syn-rand}. In these networks, non-community patterns dominate or mix with other inter-group interactions --- structures which {\sc Cesna} cannot model. On the other hand, for {\tt syn-comm}, our improvement is marginal as $K$ increases since both methods are able to handle community structures (corresponding to an $\matr{R}$ with only diagonal nonzero elements).

We also evaluate the impact of weight $\alpha$ which controls the tradeoff between link and attribute reconstruction qualities. The log-likelihoods at different $\alpha$ for methods \algrole and {\sc Cesna} are shown in Figure~\ref{fig:exp-alpha} ({\sc Censa} also has $\alpha$ as the weight for attribute log-likelihoods). We observe that log-likelihoods peaks at $\alpha=0.7$ for \algrole on both networks, suggesting $0.7$ as a possible starting point to set $\alpha$ generally. On both datasets, the performance drops quickly as $\alpha$ approaches 1 with an overemphasis on attributes. On the other hand, a small $\alpha$ has less negative impact on performance. This is possibly due to the more observations ($N^2$) for links than for attributes ($NL$ observations) so that the decrease in attribute likelihoods less affects the overall. A higher weight $\alpha$ for attributes would impact more on the overall likelihood and make the attributes better expressed.
\begin{figure}[!t]
	\centering
	\subfigure[Real-world networks]{		
		\includegraphics[width=0.5\textwidth]{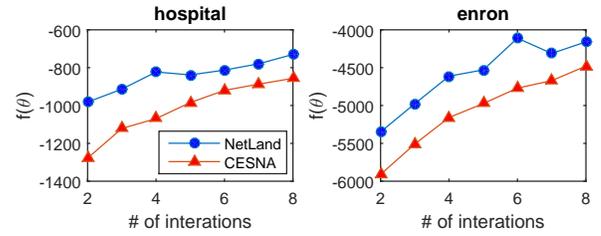}
		\label{fig:exp-ll-real}
	}
	\subfigure[Synthetic networks]{
		\includegraphics[width=0.5\textwidth]{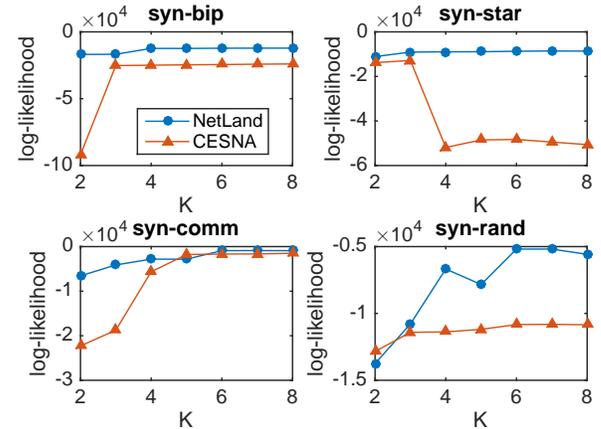}
		\label{fig:exp-ll-syn}		
	}
	\vspace{-0.5cm}
	\caption{Expressiveness: log-likelihood of network at varying $K$ using {\sc Cesna} and \algrole on real-world and synthetic datasets.}
	\vspace{-0.3cm}
	\label{fig:exp-ll}
\end{figure}

\begin{figure}[!t]
	\centering
	\includegraphics[width=0.5\textwidth]{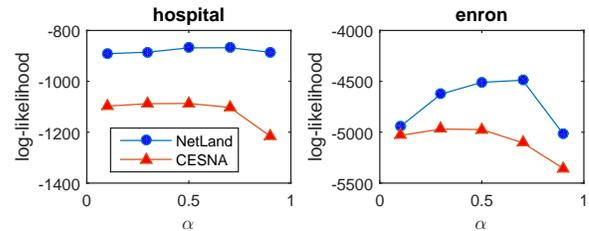}
	\vspace{-0.7cm}
	\caption{Impact of $\alpha$ on log-likelihoods}
	\label{fig:exp-alpha}
	\vspace{-0.3cm}
\end{figure}

\inlheader{Homogeneity}
The key property our method must satisfy is the homogeneity of links and attributes among the nodes in each landmark. We evaluate the homogeneity by the {\it entropy} of the link and attribute distributions in each landmark.

Since each node can belong to multiple landmarks, we approximately measure the entropy of the overlapping groups by first grouping the nodes by their main roles $\set{G}(k) = \{v: (\argmax_l x_{v,l})=k\}$. Let $\vec{p}_{k,i}=(f_{k,i}/n_k, 1-f_{k,i}/n_k)$ be the distribution of attribute $i$ in group $k$, where $f_{k,i}=|\{v\in \set{G}(k): a_{v,i}=1\}|$ and $n_k=|\set{G}(k)|$, and $\vec{q}_v=({f'_{v,l}/n_k})_{l=1\ldots K}$ be the link distribution of node $v$ with respect to the $K$ roles, where $f'_{v,l}=|\{(v,u)\in\set{E}: u\in \set{G}(l)\}|$. The mean link distribution (with respect to the roles) of group $k$ is thus $\bar{\vec{q}}_{v\in \set{G}(k)}=(\sum_{v\in \set{G}(k)} \vec{q}_v)/n_k$.

Given the grouping of nodes, we measure the homogeneity of attributes in $g(k)$ by summing up the entropy of each attribute over the nodes in $g(k)$, and measure the homogeneity of links in $g(k)$ by the summed deviation of each node's link distribution among the roles from the mean of $g(k)$ summed entropy over each attribute and links of map $\theta^*$. Finally, we give the homogeneity measures of map $\theta^*$ as follows.
\begin{align*}
h_{attrib}(\theta^*) &= \frac{1}{K} \sum_{k=1}^K\sum_{i=1}^L H(\vec{p}_{k,i})\\
h_{link}(\theta^*) &= \frac{1}{K} \sum_{k=1}^K \sum_{v\in \set{G}(k)} D_{KL}(\bar{\vec{q}}_{v\in \set{G}(k)}  \| \vec{q}_v), 
\end{align*}
where $H(\cdot)$ is (binary) entropy function and $D_{KL}(\cdot\|\cdot)$ is the KL-divergence function (relative entropy). $h_{link}$ measures how much each node's link distribution diverges from the mean distribution of its group.

We show the results for homogeneity of both links and attributes in Figure~\ref{fig:exp-entropy}. For our method, as $K$ increases, it shows an overall trend of increasing homogeneity (decreasing entropy) on both attributes and links, as expected due to unmixed roles. On the other hand, a smaller $\alpha$ generally results in higher homogeneity for attributes and slightly lower homogeneity for links, as more weight is posed on attributes in optimization.

Next, we compare our method to {\sc BinDedi} in Figure~\ref{fig:exp-entropy-vs-bindedi},  We observe that \algrole consistently outperforms {\sc BinDedi} which only considers only the topology aspect of networks. For link homogeneity, although our methods model the additional attribute observations, the resulted link homogeneity is still quite comparable to that of {\sc BinDedi} which focuses entirely on links.

We also compare our method to {\sc Cesna} which handles attributes but does not model intergroup interactions, with results given in Figure~\ref{fig:exp-entropy-vs-cesna}. Our method outperforms {\sc Cesna} in most cases by a significant margin (up to $\sim 40\%$ for {\tt enron} at $K=7,\alpha=0.5,0.9$). We explain this advantage by the greater expressiveness of our model which accounts for intergroup interactions, beyond {\sc Censa}'s state-of-the-art modeling of node-attributed networks. Moreover, we observe improvement also in attribute homogeneity over {\sc Cesna} in most cases (up to $\sim 30\%$ for {\tt hospital} at $K=7,\alpha=0.5$), as roles turns out a better concept than communities in explaining attributes.


\begin{figure}[!th]
	\subfigure[\algrole v.s {\sc BinDedi}]{
		\centering
		\includegraphics[width=0.5\textwidth]{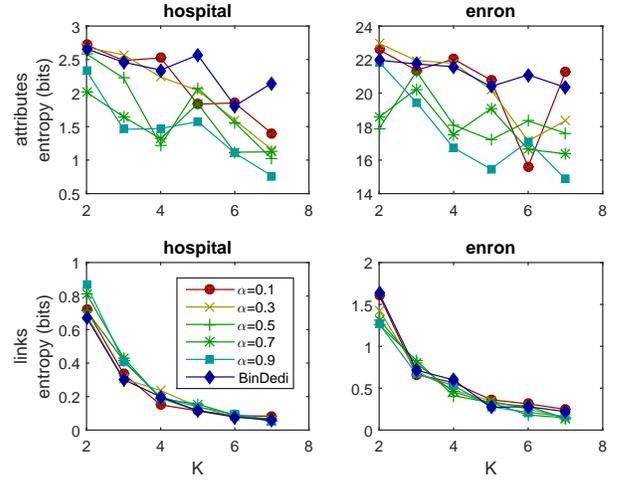}
		\label{fig:exp-entropy-vs-bindedi}	
	}
	\subfigure[\algrole v.s {\sc CESNA}]{
		\centering
		\includegraphics[width=0.5\textwidth]{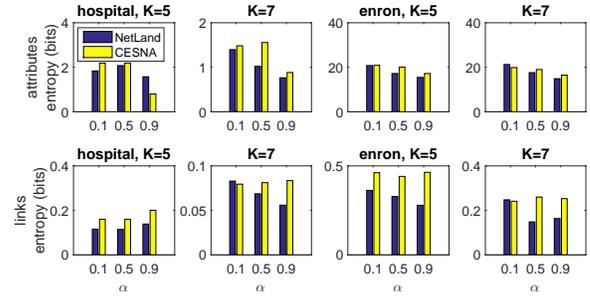}
		\label{fig:exp-entropy-vs-cesna}
	}
	\vspace{-0.5cm}
	\caption{Homogeneity: average attribute entropy $H_{attrib}$ of landmarks and link entropy $H_{link}$ of nodes.}
	\vspace{-0.3cm}
	\label{fig:exp-entropy}
\end{figure}

\begin{figure*}[!th]
	\centering
	\includegraphics[height=0.8\linewidth,angle=-90,origin=c]{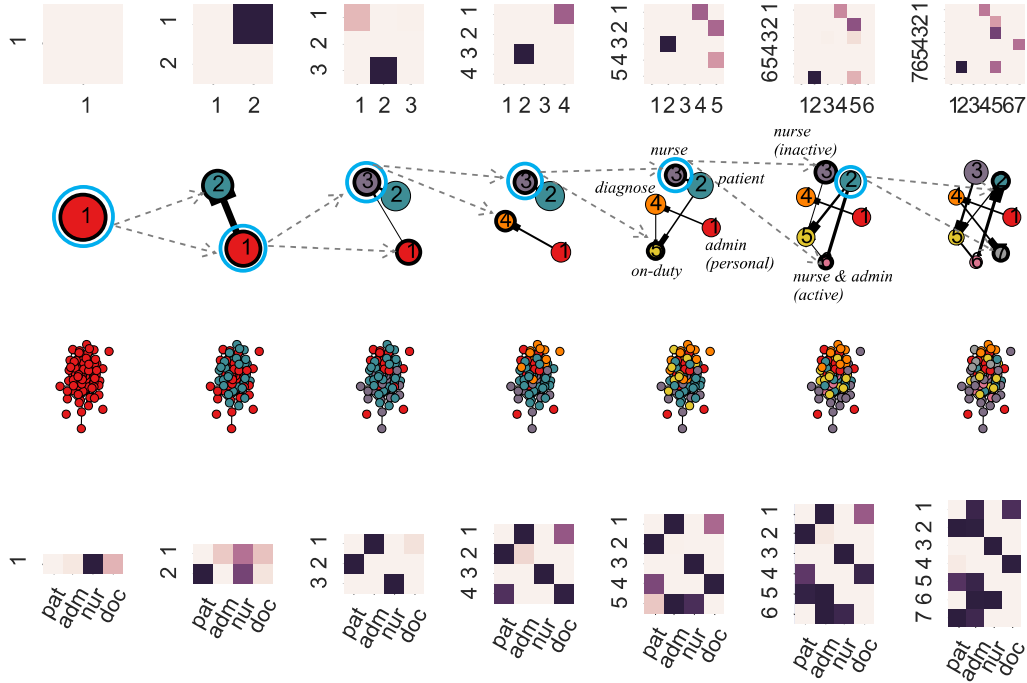}
	\vspace{-2.5cm}
	\caption{Cartography of the directed {\tt hospital} network (enlarge for details): a continuous sequence of maps $\set{M}_K$ for $K=1\ldots 7$ with increasing levels of resolution. The rows from top to bottom are: $\matr{P}_{link}$ of inter-landmark interaction probabilities, the map (landmarks selected for zooming are circled in bold blue and created sub-roles in bold black), input network (colored by main role), and $\matr{P}_{attrib}$ of probability of each attribute for each landmark. Landmarks in $\set{M}_5$ and $\set{M}_6$ are labeled by hypothesized names.}
	\label{fig:hospital-zoom}
\end{figure*}

\begin{figure}[th]
	\centering
	\includegraphics[width=\linewidth]{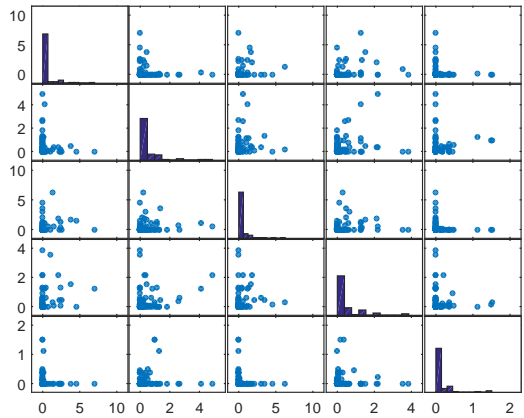}	
	\vspace{-1cm}
	\caption{Locations of nodes: coordinates of nodes ($\vec{x}_v$ for node $v$) with respect to  map $\set{M}_5$ in Figure~\ref{fig:hospital-zoom}, in the space defined by each pair of landmarks. In each subplot, the set of points represent the set of all nodes.}
	\label{fig:scatter-hospital}
	\vspace{-0.5cm}
\end{figure}

\subsection{Case Study}
In Figure~\ref{fig:hospital-zoom}, we show a sequence of maps $\set{M}_K$ by Algorithm~\ref{alg:zoom} computed by Algorithm~\ref{alg:zoom} with $\beta=0.002$ for the {\tt hospital} network. While we allow users to select any landmark to split, for this example, we split the largest role (in term of how many nodes have it as the main role) in each map. We used $c=1$ for $\matr{P}_{link}$ and $\matr{P}_{attrib}$, as defined in Section~\ref{sec:map}. For reference, we index the roles as $R_1,\ldots,R_6$.

\inlheader{Multi-resolution maps}
First of all, we can intuitively see the localized incremental changes between consecutive maps from the coloring of the non-split landmarks and the heatmaps of $\matr{P}_{link}$, and $\matr{P}_{attrib}$. The interactions between non-split landmarks are not disturbed by the zooming, too.

To start, we see in map $\set{M}_1$ that nurses has the largest population in the hospital (see corresponding $\matr{P}_{attrib}$). In the subsequent map $\set{M}_2$, role $R_2$ is characterized by mainly involving nurses and $R_1$ by patients. With help of $\matr{P}_{link}$ indicating strong interaction from $R_2$ to $R_1$, we see that the major interactions in this hospital are those from nurses to patients (possibly explained by that nurses actively perform regular checks for patients).

Increasing $K$ reveals finer-resolution structures. In the map $\set{M}_5$, role $R_1$ are administrative staff who actively approach $R_4$ which involves mainly doctors and some patients; $R_1$ possibly corresponds to ``administrative staff members assigned with specific doctors or patients to assist''. $R_2$ mainly involves patients and $R_3$ mainly nurses. $R_5$ involves mainly administrative staff and some other nurses who are passively approached by patients and doctors; they are possibly ``on-duty staff''.

To study $R_3$ in map $\set{M}_5$ in more details, we computed map $\set{M}_6$ by splitting the $R_3$ of $\set{M}_5$ into the sub-roles $R'_3$ and $R_6$. Role $R'_3$ still corresponds to nurses but now those inactive ones, $R_6$ captures active nurses, and now also some administrative staff absent in $R_3$. This is because while we enforce stability of affiliations of nodes to the non-selected roles, our formulation still allows a small number of nodes to change their role compositions (controlled by $\beta$ in \eqref{eq:obj-zoom}); particularly, some administrative staff become affiliated to the new $R_6$ in $\set{M}_6$, resulting in the added {\tt adm} attribute of $R_6$. We interpret $R_6$ as ``active hospital staff (in approaching patients)''.


\inlheader{Locatability}
Recall that each node $v$ can be located using affiliation vector $\vec{x}_v$. In Figure~\ref{fig:scatter-hospital}, we show a pairwise plot of $\matr{X}$'s columns for map $\set{M}_5$ in Figure~\ref{fig:hospital-zoom}. Each subplot shows the overall distribution of nodes in the corresponding landmarks. For instance, we observe in subplot-(5,1) that no nodes simultaneously takes roles $R_1$ (personal admins) and $R_5$ (on-calls), probably because a personal assistant does not often perform general on-call duties.

\section{Conclusion}
The task of extracting a big picture from complex real-world networks is challenging. In this paper, we presented effective cartographic methods to help with this task. We created intuitively understandable visual maps. To extract landmarks of the map, we recognized roles as the appropriate concept to model landmarks, and proposed a novel expressive model for role discovery. Furthermore, we allow for local multiple-resolution views for knowledge acquisition from networks at different granularities. We also propose efficient algorithms for computing the maps. In addition to an intuitive representation, experiments show our method outperforms state-of-the-art network models in representation accuracy by up to 27\% for real networks and up to 10 times on synthetic networks. Case study also shows our created maps actually help users to quickly capture the big picture of a network.
\label{sec:conclusion}

\bibliographystyle{abbrv}
\bibliography{ref}

\end{document}